\documentclass[preprint,11pt]{elsarticle}
\usepackage{amsfonts,amscd,amsmath, amssymb,graphicx,color}

\usepackage{amssymb}

\journal{Nuclear Physics B}

\begin{document}


\title{Towards a Quantum Theory of Solitons}

\author[a,b,c]{Gia Dvali}
\author[a,d]{Cesar Gomez}
\author[a,b]{Lukas Gruending}
\author[a,b]{Tehseen Rug}

\address[a]{Arnold Sommerfeld Center for Theoretical Physics, 
	LMU-M\"unchen, Theresienstrasse 37, 80333 M\"unchen, Germany}
\address[b]{Max-Planck-Institut f\"ur Physik,
	F\"ohringer Ring 6, 80805 M\"unchen, Germany}
\address[c]{Center for Cosmology and Particle Physics, Department of Physics, New York University
	4 Washington Place, New York, NY 10003, USA}
\address[d]{Instituto de F\'{i}sica Te\'{o}rica UAM--CSICm C--XVI
	Universidad Aut\'{o}noma de Madrid, Cantoblanco, 28049 Madrid, Spain}

	\begin{abstract}
	We formulate a quantum coherent state picture for topological and non-topological solitons.  We recognize that the topological charge arises from the infinite occupation number of zero momentum 
	quanta flowing in one direction. Thus, the Noether charge of microscopic constituents gives rise to a topological charge in the macroscopic description.  This fact explains the conservation of topological charge from the basic properties of coherent states. It also shows that no such conservation exists for non-topological solitons, which have finite 
	mean occupation number.  Consequently,  they can have an exponentially-small but  non-zero overlap with the vacuum, leading to vacuum instability. This amplitude can be interpreted as a coherent state description of false vacuum decay. 
	Next we show that  we can represent topological solitons as a convolution of two sectors that carry information about 
	topology and energy separately, which makes their difference very transparent. 	 
 Finally, we show how interaction among the solitons can be understood from basic properties of quantum coherent states.

	\end{abstract}
	\maketitle

\section{Introduction}

From the very early days of soliton physics it has been suspected that quantum solitons should be thought of as some sort of quantum coherent states. As quantum coherent states they are many quanta systems with an average number of quanta $N$ that scales as the inverse of the relevant coupling. Therefore, in the strong coupling regime this picture of the soliton as a quantum coherent state looses its meaning and we are forced to think of the quantum soliton as a fundamental particle of a potential dual theory. This is, for instance, the case in two dimensions, where Thirring fermions are the strong coupling version of Sine Gordon solitons \cite{Coleman} \cite{Mandelstan}. In this case what we identify as topological charge in the weak coupling, becomes simply fermion number in the strong coupling. 

An obvious difficulty in identifying the quantum soliton state lies in understanding the quantum meaning of the topological charge that is normally introduced in purely classical terms. In the weak coupling, where we expect to have a good quantum coherent state representation of the soliton, it is {\it a priori}  not clear how to distinguish the quanta
accounting for the topological charge from those that account for the energy. 

Intuitively it is clear that, if solitons were to admit some quantum coherent state description, 
  the dominant contribution to the energy should come from the constituents of the wavelength comparable to the Compton wavelength ($m^{-1}$) of the quanta of the theory, with their mean occupation number scaling as inverse of the coupling. 
   At the same time, we should expect that a very different type of quanta is responsible for the topology.

 In the present paper, we shall clarify this distinction by developing a coherent state picture of topological and non-topological solitons in $1+1$-dimensional theory.   We shall follow the general strategy outlined in \cite{us}, in which 
 the coherent state is constructed by using the classical Fourier-expansion data of the soliton.   
 However, we shall carefully identify and separate  topological and energetic ingredients. 
 
 We show that the contribution to the soliton mass indeed comes 
 from the quanta of  wavelength $\sim m^{-1}$ with the mean occupation number given by the inverse coupling. 
 At the same time, we observe that the origin of the topological charge is dramatically different. 
 Namely, it comes from the {\it infinite}  wavelength quanta with the net momentum flow in one direction.
 The occupation number of these topology-carriers is {\it infinite}. In other words, the mean-occupation number   
 of the coherent state constituents, $N_k$, as a function of momentum $k$ exhibits a pole in zero momentum limit, 
 $N_{k\rightarrow 0} \sim 1/k$.   This divergence in the occupation number of infinite wavelength quanta with net 
 momentum flow 
  is the quantum origin of topology.  We can think of it as a  {\it Bose sea} of such quanta. 
  
  The conservation of the topological charge then trivially follows from the basic properties of coherent states. 
  Namely, from the property that the projection of an infinite occupation number coherent state 
  on any finite-number state vanishes.  
  Due to this, the topological soliton has zero overlap with any state from the topologically-trivial vacuum sector, 
  since such states have no singularities in net occupation numbers.  
 
  This divergence is absent for non-topological solitons, and correspondingly their overlap with the vacuum 
  is non-zero, but exponentially small.  This overlap can be viewed as a coherent state description of the false vacuum instability.

 In order to make the distinction between the topological and energetic constituents, it is useful to describe the soliton 
 as a {\it convolution} of the two sectors, i.e.,   
 \begin{center}
soliton $=$ ( topology ) $\star$  ( energy ) \, .
\end{center}
This way of thinking allows to represent a soliton coherent state as a tensor product of two states,
$|sol\rangle = |t\rangle \otimes |E\rangle$, one  ($ |t\rangle$) describing the topology and the other 
($ |E\rangle$) the energy. This representation allows to directly observe that the key difference 
between topological and non-topological solitons is encoded in the nature of the $|t\rangle$-state.  

 We perform different consistency checks, and show that certain known properties of solitons, e.g., the 
 exponentially weak interaction among separated solitons can be nicely understood from basic properties of coherent states.  
 
 Although we limit our analysis to $1+1$-dimensional theory, the basic results of our construction can be easily generalized to solitons in arbitrary number of dimensions. 
 These generalizations show that the emergence of topological charge from the divergence in occupation number 
 of infinite wavelength quanta with net momentum (or angular momentum) flow is an universal property.

\section{Coherent State Picture of Non-Topological Soliton} 

 In order to clearly identify the role of topology, we confront corpuscular resolutions of non-topological and 
 topological solitons.  We start with non-topological solitons.

 Consider a $1+1$-dimensional  theory of a classical field $\phi(x_{\mu})$  ($x_0=t, x_1=x$) with the Lagrangian 
 \begin{equation}
   {\mathcal L} \, = \,  (\partial_{\mu}\phi)^2  \, - \, m^2\phi^2 \, + \, g^2 \phi^4 \, , 
 \label{01}
 \end{equation}
   where $m^2, g^2 \, > \, 0$.  This theory has a classically-stable vacuum at $\phi \, = \, 0$, but becomes unstable for large field values.   Correspondingly, there is a static soliton solution which describes a local (in $x$) excursion of the field, from the point $\phi=0$ to the point  $\phi = m/g$ and back. 
     
  If we think of the coordinate $x$ as time, the soliton configuration describes a classical evolution of the field in the inverted potential. 
    The field starts at $\phi = 0$ at $x=-\infty$,  reaches the point $\phi = m/g$ at $x=0$ and bounces back 
    reaching the point $\phi=0$ at $x= + \infty$. The soliton solution is given by
    \begin{equation}
    \phi_{c}(x) \, = \, {m\over g} {\rm sech} (mx) \, , 
    \label{sol01}
    \end{equation}
  which, as said above, describes an  excursion of the field  across the barrier and back.  The energy 
  of the configuration is 
  \begin{equation}
  E_{non-top}  \, = \, {4 m^3 \over 3g^2}.    
    \label{energy1}
   \end{equation}

 Expanding the 
 classical solitonic field in a plane wave basis ($2\pi R$ is the regularized volume of space),  
   \begin{equation}
  \phi_{c}(x) \, = \,  \sqrt{R}  \int  \, {d k \over \sqrt{4\pi |k|}}  \, ( e^{ikx}   \alpha_k \, + \,  e^{-ikx}   \alpha_k^*) \,,   
 \label{Fieldoperator}
 \end{equation}
  the $\alpha_k$ and $\alpha_k^*$ are c-number functions of momentum $k$ that satisfy 
    \begin{equation}
  \alpha_k^*\alpha_k  \, = \pi { |k| \over R} {1 \over g^2} {\rm sech}^2 \left ({\pi k \over 2m} \right ) \,. 
  \label{acoeff}
  \end{equation}  
    Using this representation, the classical  energy of the soliton takes the form, 
   \begin{equation}
  E_{non-top}  \, = \,  \int_k  |k| \alpha_k^* \alpha_k  \, ,  
 \label{Mclass}
 \end{equation}  
 where $\int_k \equiv R \int dk$. \\

We now wish to represent the soliton as a quantum coherent state, $|sol\rangle$.  The natural way of doing this 
is to identify the Fourier expansion coefficients of the classical field, $\alpha_k$, with the expectation values 
of Fock space operators  
 $\hat{a}_k, \hat{a}^{\dagger}_{k}$, 
 \begin{equation}
 \langle sol|\hat{a}_k |sol\rangle \, = \, \alpha_k \,, 
 \label{expalpha}
 \end{equation}
  satisfying the creation-annihilation algebra
\begin{equation}
\label{algebra1}
[\hat{a}_k,\hat{a}^{\dagger}_{k'}] = \delta_{k,k'} \, ,
\end{equation}
for momenta $k,k'$.
 
 The operators $\hat{a}^{\dagger}_{k}, \hat{a}_k$ should not be confused with 
creation and annihilation operators of asymptotic propagating quanta of the theory. Rather, they create and destroy  solitonic
constituents, which have very different dispersion relation as compared to the free quanta.
A more detailed discussion on the nature of these constituents will be given below in section four.

It then follows, from the general  properties of coherent states,  that $|sol\rangle$ must be a tensor product 
of coherent states for different $k$, 
 \begin{eqnarray}
	\label{cohsol1}
	  |sol\rangle \  = \prod_{\otimes k} | \alpha_k \rangle \, , 
 \end{eqnarray}	
 where $| \alpha_k \rangle $ is a coherent state for momentum mode $k$, 
 \begin{eqnarray}
	\label{cohsol}
	|\alpha_k \rangle =  e^{-{1\over 2} |\alpha_k|^2} e^{ \alpha_k \hat{a}^{\dagger}} |0\rangle\, = \,  
	e^{-{1\over 2} |\alpha_k|^2} \sum_{n_k=0}^{\infty}  \frac{\alpha_k^{n_k}}{\sqrt{n_k !}} | n_k \rangle \, .
\end{eqnarray}	
Here  $|n_k \rangle$ are number eigenstates of corpuscles with momentum $k$. 

 Defining the particle number operator in the standard way, $\hat{N}_k \, \equiv \, \hat{a}_k^{\dagger}\hat{a}_k $, it is clear that 
 the quantity,  
  $N_k \equiv  \langle sol|\hat{a}_k^{\dagger}\hat{a}_k  |sol\rangle \,  =   \alpha_k^*\alpha_k$ counts the mean occupation 
 number of corpuscles of momentum $k$ in the state $|sol\rangle$. Correspondingly, 
 the quantity 
  \begin{equation}
  N \, \equiv \, \int_k \, N_k  \, = \, \int_k \, \alpha_k^*\alpha_k
  \label{Ntotal}
  \end{equation}
  is the total mean occupation number. 
Using (\ref{acoeff}) and (\ref{Mclass}), we find that the total mean occupation number of corpuscles
and the energy of the soliton are given by
   \begin{equation}
  N \, = \, \int_k \, N_k \, = \, {m^2 \over g^2} \, \left ( {{\rm log}(2) 8 \over  \pi} \right )  
  \label{Nnontop}
  \end{equation}
  and
   \begin{equation}
    E_{non-top} \, = \, \int_k |k| N_k \, = \, {4 m^3 \over 3g^2} \, , 
  \label{massink}
  \end{equation}
  respectively.

  Thus,  the dominant contribution  both to the number $N$ as  well as to the mass of the soliton
  comes from the corpuscles of the momentum $k \lesssim m$. They also set the size of the soliton 
  as $\sim m^{-1}$.  
  The fact that the number  (\ref{Nnontop}) is finite for non-topological solitons has very important 
  consequences for the quantum description.  This finiteness is the reflection of the fact that 
  the non-topological soliton carries no conserved quantum number. Correspondingly, the vacuum is not protected against the instability of soliton-creation! 
    In fact, this instability reflects the fact that the vacuum $\phi=0$ is not a true vacuum of the theory and decays via 
    bubble-nucleation \cite{ColemanII}. 
   We shall discuss this in the next section.

  \subsection{Consistency  Check: Quantum Nucleation of Non-Topological Soliton and Vacuum Decay}

    Let us evaluate the amplitude of quantum fluctuations that creates a soliton out of the vacuum state. 
  This amplitude is given by the overlap between the vacuum $|0 \rangle$   and the soliton $|sol \rangle$. 
    In the coherent-state picture  this overlap is equal to   
  \begin{equation}
 \langle 0| sol \rangle \, = \, e^{- {N \over 2}} \, . 
 \label{overlap0N}
 \end{equation}  
    Notice, that this expression is universal for any coherent state. The difference is only in the number $N$.   
   
   For a non-topological soliton,  using (\ref{Nnontop}) and restoring powers of   
    $\hbar$, we obtain the following scaling of the exponential
 amplitude   
      \begin{equation}
 \langle 0| sol \rangle \, = \, e^{- {m^2 \over \hbar g^2}} \, ,  
 \label{overlap0N1}
 \end{equation}   
 where we have ignored the order one numerical coefficient of (\ref{Nnontop}). 
 This amplitude is zero only in the classical limit, in which $\hbar \rightarrow 0$ and $m, g$ are kept finite\footnote{In our notations, 
 $m$ is the frequency of a classical field oscillation and it is non-vanishing in the limit $\hbar=0$. 
 The mass of a quantum excitation (energy of a one-particle state) is $\hbar m$ which consistently vanishes in the classical limit.}. 
 
 This result has a very interesting physical meaning and  amounts to the fact that quantum-mechanically the vacuum $\phi =0$ 
 is unstable.  In this respect, the possibility of non-topological soliton-creation reflects the instability of the $\phi=0$ vacuum at the quantum level. 
 
    Notice, that in the semi-classical limit, in which the classical field configuration is not resolved as a coherent state,  the decay  
 goes through the nucleation of a critical bubble which must have the same classical energy as the  vacuum $\phi=0$, and thus, cannot be static. This ``bubble" is a configuration that interpolates  between $\phi=0$ and  $\phi=\phi_* > m/g$ in order to have the zero total  energy.   
  But quantum-mechanically the decay can go through the intermediate creation of other coherent states, such as 
  the above static soliton,  since none of them is a true energy eigenstate\footnote{Notice, that the channel of decay through static soliton nucleation is not necessarily a new channel, since we never proved that this coherent state and 
  the one corresponding to the critical bubble of zero classical energy are orthogonal.  To resolve the latter as a coherent state is a bit more complicated because of the lack of the exact classical solution.  It would be interesting to study this issue further.} .   

 Only in  the classical limit ($N \rightarrow \infty$), the non-topological soliton becomes an energy eigenstate and correspondingly the amplitude (\ref{overlap0N}) vanishes in this limit.  This is a very important result of how 
 the coherent state picture tells us about the instability of the vacuum. In other words, the moment we declare that the static soliton exists and is represented as a coherent state of finite occupation number, it can be nucleated out of the vacuum and lead to its instability.

 \section{Coherent State Picture of Topological Soliton}
 
   The next example we want to consider is a soliton 
 with non-zero topological charge.  For this we have to slightly modify the Lagrangian 
 (\ref{01}) in order to  create two degenerate stable minima.  The new Lagrangian reads 
  \begin{equation}
   {\mathcal L} \, = \,  (\partial_{\mu}\phi)^2  \, - \, g^2(\phi^2 \, - \, m^2/g^2)^2\, . 
 \label{topL}
 \end{equation}
 This theory has two degenerate minima at $\phi \, = \, \pm m/g$ and there is a  solution 
 that connects the two.  Again, this soliton corresponds to the motion of a particle in an inverted
 potential in time $x$. The particle starts at  $\phi \, = \, \pm m/g$ at $x= -\infty$ and reaches $\phi \, = \, \mp m/g$
at $x= +\infty$. The solution is described by a kink or an anti-kink,   
  \begin{equation}
 \phi_{c}(x) \, = \, \pm \, {m \over g}  {\rm tanh}  (xm) \, .  
 \label{kink}
 \end{equation}
 
  Repeating exactly the same construction as in the case of the non-topological soliton, we shall now represent the kink as a coherent state (\ref{cohsol1}) with the occupation number data 
 $N_k$ fixed by matching it with the coefficients of the Fourier expansion of the soliton field 
 (\ref{Fieldoperator}). 
 For the topological soliton (\ref{kink})  this matching gives
 \begin{equation}
  N_k \, = \, \alpha_{k}^* \alpha_{k} \, = \pi { |k| \over R} {1 \over g^2} {\rm csch}^2 \left ({\pi k \over 2m} \right ) \,. 
  \label{coefficient1}
  \end{equation}
  The mass of the soliton is correspondingly 
  \begin{equation}
    E_{top} \, = \, \int_k |k| N_k \, = \, {8 m^3 \over 3g^2} \, .  
  \label{masssoltop}
  \end{equation}
 
 Notice, that unlike the non-topological case, the occupation number $N_k$ exhibits a $1/k$-type singularity for small $k$.    
 This singularity is very important and is the quantum mechanical manifestation of the topological charge.
  Due to it, in case of the topological soliton, the total number  $N$   exhibits a logarithmic  
 divergence for small $k$, 
  \begin{equation}
N \, = \,  \int_{k_0} dk N_k \, \sim \, {\rm log}(k_0)|_{k_o\rightarrow 0}  \rightarrow  \infty \, . 
\label{divergence} 
\end{equation} 
Notice, that the quanta with  $k\rightarrow 0$ contributing to this divergence do not contribute to the energy of the soliton  
  (\ref{masssoltop}) to which
 the contribution from small $k$-s vanishes. The infrared quanta only contribute to the topological charge. 
 This  divergence ensures that topological charge 
 is conserved in a full quantum theory.  

In particular, creation of a topological  soliton out of the vacuum, unlike the non-topological one,  is impossible in the quantum theory.  
The coherent state corresponding to this soliton, even for finite $\hbar$, has {\it infinite}  occupation number of zero momentum modes.  Correspondingly, the  soliton-creation amplitude 
  vanishes already at the quantum level. 
To see this it is enough to substitute in (\ref{overlap0N}) the expression  (\ref{divergence}). We get 
      \begin{equation}
 \langle 0| sol \rangle \, = \, e^{- {1 \over \hbar} \infty} \, . 
\label{overlap0N2}
 \end{equation}     
    Although the zero momentum modes  give vanishing contribution 
  to the soliton mass (\ref{masssoltop}), they play the crucial role for the  conservation of the topological 
  charge, as it is clear from (\ref{overlap0N2}).  
  
      Thus, the following quantum picture emerges.  Classically,  the topological charge of the soliton comes from the boundary 
 behaviour of the field which asymptotes to different constant values for $x \, =\,  \pm \infty$. 
Correspondingly, in quantum description   this charge is determined entirely by the  infinite-wavelength corpuscles. 
The occupation  number of $k=0$ modes is infinite which ensures that topological charge is {\it not}  subject to quantum fluctuations.   

In contrast, the mass of the soliton gets the dominant contribution from 
$k \lesssim {\hbar m} $ modes.  The mean occupation number of  such modes is finite ($ N_m \, \sim \,  {m^2 \over \hbar g^2}$) \footnote{We mean that the  contribution of the modes with momentum $0 \leqslant k < \epsilon$ to 
the mass of the soliton is  
 $\sim \epsilon$ and is negligible for $\epsilon \ll m$. Whereas, the contribution of the same modes to the topological charge is dominant no matter how small $\epsilon$ is.}.  Because of their finite number the contribution of these modes to physical processes is in general subject to quantum fluctuations. 
  Consequently, the quantities that are determined classically by the mass of the soliton are expected to be subjected to quantum corrections in the quantum picture. 
  
  \section{Corpuscular Creation and Annihilation Operator Algebra}
  As advertised in section 2, we will now clearly define what we mean by corpuscular quanta and why these are
  completely different from perturbative $S$-matrix states. For that purpose, we shall present the basic steps leading to the identification of the corpuscular creation and annihilation operator algebra (see equation (\ref{algebra1})) on the basis of which we define the coherent state quantum portrait of the classical solitons. In order to do that it would be convenient, following Coleman \cite{Coleman},  to work in quantum field theory in the Schr\"odinger picture. Notice, however, that in explicit applications we do not need
  this particular construction. Rather it can be viewed as a formal way of defining a conjugated momentum in the  Schr\"odinger picture based on the algebra of operators.

Due to locality the set of commuting observables are $\hat \phi (x)$ where as before $x$ denotes the spatial coordinate. Let us introduce a generic function $\omega(k)$ that we shall not fix for the time being and a formal algebra of annihilation and creation operators $\hat a_k, \hat a^{\dagger}_k$ with the usual commutation relations associated with this $\omega(k)$. The operator $\hat a^{\dagger}_k$ creates quanta with momentum  $k$ relative to a vacuum state $|0\rangle$ defined by $a_k|0\rangle =0$. We shall refer to  this algebra as {\it the corpuscular algebra} and to the corresponding Fock space $F_{\omega(k)}$ as {\it the corpuscular Fock space}. 

Let us now define the quantum field operators $\hat \phi(x)$ as
\begin{equation}
\hat \phi(x) = \sqrt{R}\int d k \frac{1}{\sqrt{4 \pi\omega(k)}} (\hat a_k e^{-ikx} + \hat a^{\dagger}_k e^{ikx}) \,.
\end{equation}
Note, that we can formally define a canonical momentum in terms of the creation and annihilation operators as follows:
\begin{equation}
\hat \pi(x) =i \frac{1}{\sqrt{2 \pi R}}\int d k \sqrt{2\omega(k)} (\hat a_k e^{-ikx} - \hat a^{\dagger}_k e^{ikx}).
\end{equation}
The dynamics, i.e. the time evolution, is introduced once we identify  the Hamiltonian operator. In particular, for a  generic state $|s\rangle$ we have  
\begin{equation}
\phi_{s}(x,t) = \langle s (t)| \hat \phi(x) |s(t) \rangle \,, 
\end{equation}
where $|s(t)\rangle$ is defined by $i\partial_t |s(t)\rangle =H |s(t)\rangle$
for $H$ being the Hamiltonian of the system. 

In contrast, if we are working perturbatively around a classical non-solitonic vacuum we define the free Hamiltonian $H_0$ relative to the notion of asymptotic $S$-matrix particles and we introduce creation and  annihilation operators $\hat b^{\dagger}_k,\hat b_k$ for these particles. In terms of these operators, 
$\hat H_0 = \omega_0(k) \hat b^{\dagger}_k\hat b_k $, and the full Hamiltonian is  $\hat H = \hat H_0 +$interaction terms.  
The $\omega_0(k)$ is fixed by the dispersion relation for the asymptotic $S$-matrix particles. 
In order to write the former quantum operator $\hat \phi(x)$ in terms of the $S$-matrix algebra of $\hat b_k,\hat b^{\dagger}_k$ we need to relate $\hat b_k, \hat b^{\dagger}_k$, i.e., the asymptotic $S$-matrix algebra, and the {\it corpuscular algebra} $\hat a_k,\hat a^{\dagger}_k$.

Before doing this let us use the corpuscular algebra to define a coherent state $|sol\rangle$, such that
\begin{equation}
\langle sol| \hat \phi(x) |sol \rangle = \phi_c(x).
\end{equation}
This is what we have done in the previous sections and this procedure effectively reduces to defining the coherent state in terms of the Fourier components of the classical configuration as follows,
\begin{equation}
\hat a_k|sol\rangle =\alpha_k |sol\rangle.
\end{equation}
Here the $\alpha_k$ are again the Fourier components. As a coherent state, the energy of this state is simply given by
\begin{equation}
E = R\int  dk \omega(k) N_k,
\end{equation}
where $N_k = |\alpha_k|^2$. Matching with the classical energy of the BPS solution, $E=\int dx (\partial \phi_c(x))^2$,
we obtain $\omega(k)=|k|$ as the dispersion of the corpuscles\footnote{
Note, that this is twice the usual kinetic part of the energy. This, of course, does not correspond to a free Hamiltonian, but is rather
a consequence of the non-trivial relation between kinetic and potential energy on the BPS equation. }.

Note that this {\it corpuscular dispersion relation} is generically very different from the dispersion relation of the $S$-matrix free particles. This makes explicit that the corpuscles in terms of which we define the soliton as a coherent state and the $S$-matrix particles are very different entities. In particular, the corpuscles do not define asymptotic $S$-matrix states.

 The situation is crudely analogous to defining quarks as constituents of a baryon in  QCD with $N_c$-colors. 
 The constituents have the wavelengths of the size of baryon (i.e., inverse QCD scale), but obviously, their dispersion relations have nothing to do with the ones of the asymptotic $S$-matrix states.  
  
Now we can try to unveil the relation between the corpuscular algebra and the $S$-matrix algebra. The relation is given by the BPS condition on the soliton coherent state,
\begin{equation}\label{one}
\langle sol|\hat H(\hat b_k,\hat b^{\dagger}_k)|sol \rangle = \langle sol|\int dk \omega(k) \hat a^{\dagger}_k\hat a_k  |sol \rangle \, .
\end{equation}
Of course, this relation is  a highly  non-linear relation between the $S$-matrix quanta $\hat b_k,\hat b^{\dagger}_k$ and the {\it corpuscles} $\hat a_k,\hat a^{\dagger}_k$. In particular, notice that the 
vacuum of the corpuscular algebra is {\it  not } the $S$-matrix vacuum. 

 In other words, the free $S$-matrix quanta $\hat b_k,\hat b^{\dagger}_k$  and the solitonic corpuscles  $\hat a_k,\hat a^{\dagger}_k$
 diagonalize the Hamiltonian on two very different states, the perturbative vacuum and the soliton-state, respectively.

Only on the soliton state we get the basic relation (\ref{one}). Note also that the corpuscular algebra is defined in terms of the particular soliton we want to describe.
We shall have to say a couple more words on the precise nature of these corpuscles in the next section.
    
 \section{Topology-Energy decomposition: Topological Soliton as Convolution}
 
  The diverse roles of  the quanta contributing into the energy and topology can be nicely visualized 
 in a picture in which the soliton is defined as a {\it convolution}  
 between topology and energy, i.e.,
\begin{center}
soliton $=$ ( topology ) $\star$  ( energy ) \, .
\end{center}
The quantum mechanical meaning of the above decomposition is based on the fundamental theorem of convolution, namely that the Fourier-transform -- which is the tool we need in order to promote classical $c$-numbers into creation and annihilation operators -- of the convolution of two functions is just the product of the Fourier-transforms. This will convert the former classical relation into
\begin{center}
$|sol\rangle$ $=$  $|t\rangle$ $\otimes$ $|E\rangle$,
\end{center}
where $| t \rangle$ and $| E\rangle$ are coherent states constructed out of quanta carrying information about topology and energy, respectively.
 
To illustrate this program explicitly, consider the kink solution (\ref{kink}) .  In order to disentangle the topology from the energy at the classical level we use the relation
 \begin{equation}
{\rm sign}(x)\star f(x) = + \frac{1}{2} \int_{-\infty}^x f(x')dx' - \frac{1}{2} \int_x^{\infty} f(x') dx'\,, 
\end{equation}
with ${\rm sign}(x)$ being the sign function. Using this expression, we easily get
\begin{equation}
\phi_c(x) =  \frac{m}{g} \Big({\rm sign} \star \operatorname{sech}^2 \Big) (mx) \,, 
\end{equation}
where the sign part represents the topology, whereas the $\operatorname{sech}^2$ represents the energy component. It is important to note already at this level that the energy component has the profile of the square of the non-topological soliton  (\ref{sol01})  for the inverted potential. This tells us that this component has no knowledge about the topology.  

In order to move to quantum mechanics, we first define the Fourier-transform slightly changing the parametrization 
with respect to (\ref{Fieldoperator}),  
\begin{equation}
\phi_c(mx) = \frac{1}{2} \int \frac{d(\frac{k}{m})}{\sqrt{4\pi (\frac{k}{m})}} \Big( \alpha_k e^{i (\frac{k}{m})xm} + {\textrm{h.c.}} \Big) \, .
\end{equation}
 Using now the fundamental convolution theorem, we get
 \begin{equation}
 \alpha_k = \frac{m}{g\sqrt{\pi}} \sqrt{\frac{k}{m}}  [F({\rm sign})] (k/m) [F ({\rm sech^2})] (k/m) \, , 
 \end{equation}
 with $F$ representing the Fourier-transform,
 \begin{equation}
F({\rm{sech}^2}) = \pi \frac{k}{m} {\rm{csch}} \Big( \frac{\pi k}{2m} \Big)
\label{FE}
\end{equation}
and 
 \begin{equation}
F({\rm sign}) = \frac{im}{k} \, .
\label{FT} 
\end{equation} 
 The expressions (\ref{FE}) and (\ref{FT}) carry information about the energetic and the topological composition
 of the soliton.  
   Notice the pole at $k=0$ in 
   (\ref{FT}).  This pole encodes quantum information about the topology of the soliton.  Namely, as before, topology comes 
   from the infinite occupation number of zero-momentum corpuscles that support the momentum-flow in one direction. 
   The only novelty in the convolution picture is that the pole is clearly attributed to one side  ($t$-sector) of the decomposition. 
     On the other hand, no such divergent part appears in (\ref{FE}). The contribution to this expression mainly comes 
     from modes with momentum spread $|\Delta k|  \sim m$ and zero net momentum. This is the reflection of the fact that such modes contribute to the energy of the soliton 
     without contributing to the topology.  They do not support momentum flow in any direction.

The above results lead us to the following representation for $\alpha_k$, 
\begin{equation}
\alpha_k = t_k\,  c_k \, , 
\end{equation}
where 
\begin{equation}
c_k \, \equiv \,  \sqrt{\pi m}  \frac{k}{g} {\rm{csch}} \Big( \frac{\pi k}{2m} \Big)
\label{CK}
\end{equation}
and 
 \begin{equation}
t_k \equiv  \frac{i}{\sqrt{k}} \, .
\label{TK} 
\end{equation} 
We now define two sets of creation-annihilation operators $\hat{t}_k^{\dagger}, \hat{t}_k,$ and 
 $\hat{c}_k^{\dagger}, \hat{c}_k$, which satisfy the standard algebra, $[\hat{t}_k, \hat{t}_{k^{'}}^{\dagger}] = \delta(k-k')$ 
and $[\hat{c}_k, \hat{c}_{k^{'}}^{\dagger}] = \delta(k-k')m^2$ and all other commutators vanishing.  Next, we construct the 
corresponding coherent states, 
\begin{equation}
\hat{c}_k|c_k\rangle = c_k |c_k\rangle
\end{equation}
and
\begin{eqnarray}
\label{top}
\hat{t}_k|t\rangle = t_k |t_k\rangle\, , 
\end{eqnarray}
where $c_k$ and $t_k$ are given by (\ref{CK}) and (\ref{TK}), respectively. 
Since the operators $\hat{c}_k$ and $\hat{t}_k$ act on different Fock spaces, we 
 can write the coherent state representing the soliton in the following form:
\begin{eqnarray}
	\label{cohsol}
| sol\rangle \,  = \,  | t\rangle \otimes |E \rangle,
\end{eqnarray}	
where $| t \rangle \equiv \prod_{k} | t_k\rangle$ and $| E \rangle \equiv \prod_{k} | c_k\rangle$.



The quanta $\hat{c}_k$ are not different from the constituents of non-topological solitons built on the zero topological sector. 
 The state $|E\rangle$ is defined as a quantum coherent state relative to these quanta. 
 The state $|t\rangle$ is just the coherent state $|{\rm sign}(x)\rangle$. \\
 
 Let us pause and make a couple of comments on the meaning of the above coherent states. 
 
 Formally, as a coherent state of $t$-quanta the state  $|t\rangle$ is characterized by an infinite occupation number of zero momentum modes $\hat t_0|t\rangle = \infty |t\rangle$ and can be interpreted as a sort of {\it Bose sea}. On the other hand
note, that as a convolution the non-topological soliton can be trivially represented as
$\delta (x) \star \rm sech(x)$ and therefore we can proceed as we have done with the topological soliton and associate to  it a quantum state $|t\rangle \otimes |E\rangle$ with $|E\rangle$ exactly the type of coherent state associated to the non-topological soliton considered in the first section of the paper. This makes clear that from the energetic point of view topological and non-topological solitons are the same sort of coherent states. However, in the non-topological case the state $|t\rangle$ is trivially defined by $\hat t_k|t_{nt}\rangle = |t_{nt}\rangle$ instead of $\hat t_k |t\rangle = \frac{i}{\sqrt{k}}|t\rangle$. We can thus think about the topological charge as of the order $n$ of the pole at $t_{k=0}$. Naturally, $n=0$ holds for the non-topological soliton. Under these conditions the topological stability reduces simply to the conservation of $n$. The coherent state understanding of this conservation lies in the fact that a non-vanishing $n$ requires an infinite Bose sea of quanta. \\

 Secondly,  we have to note,  that operators defined as $\hat{a}_k= \hat{t}_k\hat{c}_k$ no longer satisfy the 
 standard creation-annihilation algebra. This a priory is not necessarily a fatal problem, since the solitonic constituents 
are not propagating asymptotic degrees of freedom  anyway. They only exist in form of the bound-state and are off-shell relative to their 
asymptotic counterparts as discussed in section 4. This is already clear from their dispersion relations.  In particular, due to time independence of the kink in $1+1$ dimensions,
 they have zero frequencies, while having non-zero momenta. 
Such a dispersion relation is only possible for the interaction eigenstates.  Among the free-particles, such a dispersion relation is exhibited by tachyons of momentum exactly equal to the absolute value of their mass.  Or equivalently, 
tachyons moving at an {\it infinite}  speed.  Thinking of the solitonic constituents as of tachyons has certain appeal in the 
following sense. First, the tachyons  cannot be the asymptotic $S$-matrix states. They cannot exist in the free state, but only 
within the bound-state in form of the soliton, similarly to quarks that can only exist inside hadrons. 
Secondly, at the same time the zero-frequency tachyons move at infinite speed. That is, they 
create an instant  momentum flow between the two spatial infinities  and this is why they ``smell" topology. 
 However, 
the particular interpretation of the nature of constituents is not affecting the physical consequences of the coherent state picture. Therefore, we 
shall not insist on any of them. 
 \\

\subsection{Orthogonality of Vacua with Different Topologies} 

 Again, the above construction automatically accounts for the fact that the soliton of topological charge
different from $0$ has vanishing overlap with any state with zero topological charge. 
Indeed, the scalar product of this coherent state with the vacuum is given by 
\begin{eqnarray}
\label{overlap1}
|\langle 0 | sol \rangle|^2\,  = \, e^{- \int dk |\alpha_k|^2}\, .
\end{eqnarray}
Note now that according to (\ref{top})  for the coherent state $| t \rangle$,  we have  $\alpha_k = t_k$,  and  because of 
(\ref{TK}) the integral  $\int dk |\alpha_k|^2 \, = \, \int dk/|k| $ diverges.   
Thus,  computing the overlap of the topologically-trivial vacuum with the topological coherent state
we get 
\begin{eqnarray}
\label{overlap}
|\langle 0 | t \rangle|^2\,  = \, e^{- \int dk t_k^* t_k} = 0\, . 
\end{eqnarray}
 Obviously, by the same reason the overlap between the topological soliton and any other state constructed about the 
topologically-trivial vacuum is also zero.
For the coherent state $|E\rangle$ the average number of energy quanta can be defined as
\begin{equation}
N= m^{-1} \int d(k/m) \langle E| \hat{c}_k^{\dagger}\hat{c}_k |E\rangle\, = \, \frac{8 m^2}{3g^2} \, , 
\end{equation}
which is of the same order as the number of quanta in the case of the non-topological soliton (\ref{Nnontop}) obtained earlier.  

\subsection{Energy and Topological Charge}

 We shall define the normalized topological number, i.e., the one representing the different classes in the $\pi_0$ of the vacuum manifold in terms of the topological {\it quanta} as
 simply the order of the pole of $t_k$ at $k=0$ or equivalently by the Cauchi principal value,
\begin{eqnarray}
\label{winding}
|Q|= \frac{1}{\pi} {\rm{Im PV}} \int dk \langle t|\hat{t}_k^{\dagger}\hat{t}_k|t\rangle \, . 
\end{eqnarray}
As explained above, the pole in $t_k$ at $k=0$ implies the existence of a Bose sea with an infinite number of zero momentum $t$ quanta and, therefore, with zero overlap with states defined by regular $t_k$ at $k=0$ \footnote{In general, we can use the convolution to define the topological charge as follows. Given a solitonic classical field configuration we look for a non trivial representation of type $f\star g$, where neither $f$ nor $g$ are delta functions and where one of the two functions, let us say $g$, is a gradient. In these conditions topology is associated with $f$. Denoting $F_k$ its Fourier coefficients the {\it topological charge} is determined by the singularity of $F_k$ at $k=0$. Moreover the role of the {\it vacuum} in the topologically non trivial Hilbert space is played by the {\it Bose sea} coherent state defined by  $F_{k=0}$.}  .

The energy is instead given by
\begin{equation}
E_s= \int d(k/m) \langle E|\hat{c}_k^{\dagger}\hat{c}_k|E\rangle \, .
\label{energy3}
\end{equation}
and it is equal to $\frac{8 m^3}{3g^2}$.
Note, that we can understand the energy of the soliton as a collective effect
of $N \, = \, \frac{8 m^2}{3g^2}$ quanta, each contributing a portion $m$ to the energy.

Since the energy only depends on the corpuscular quanta $\hat{c}_k$ we can define it as the expectation value of a corpuscular Hamiltonian $H(\hat{c}_k, \hat{c}^{\dagger}_k)$ in the coherent state $|E\rangle$. 
At leading order in $1/N$ 
 the {\it corpuscular Hamiltonian} $H$ can be approximated as
\begin{equation}
H= \sum_{k} \hat{c}^{\dagger}_k \hat{c}_k \, .
\label{hamiltonian1}
\end{equation}

  It is instructive to view the corpuscular structure of the soliton in the light of many-body physics.  A generic corpuscular Hamiltonian can schematically be visualized as 
 \begin{equation}
H \, = \, \sum_{k} \hat{c'}^{\dagger}_k \hat{c'}_k \, + \,  \, {\rm interaction~terms}  \,, 
\label{generalH}
\end{equation}
where prime indicates that in general we are taking $1/N$ effects into account.
The ``interaction terms" stands for up-to-quartic momentum-conserving contractions (e.g., $\sum_{k_1, k_2, k_3}  \hat{c'}_{k_1} \hat{c'}^\dagger_{k_2} \hat{c'}_{k_3} \hat{c'}^{\dagger}_{k_1 + k_2 - k_3} \, +\, ...$).
 Then the soliton corresponds to a state with distributions of $c_k$-numbers  over which  the Hamiltonian 
 is effectively diagonal and takes up the form (\ref{hamiltonian1}), up to $1/N$ corrections.

Finally, let us discuss the zero mode from the corpuscular point of view.  This mode corresponds to translations of the soliton profile in the $x$-direction and it is the Nambu-Goldstone mode of broken translations. 
 At the corpuscular level such a transformation simply amounts to redefining the corpuscular operators by a phase, $\hat{c}_k\rightarrow e^{i k b} \hat{c}_k$, where $b$ is the displacement of the profile along the $x$-direction. Clearly, (\ref{hamiltonian1}) is invariant under this $U(1)$ transformation, as it should be. 
 
 \subsection{Convolution of Field Operators}
 At the quantum level the former representation of the soliton configuration as a convolution of topology and energy  implies the following representation of the quantum field operator $\hat \phi(x)$,
 \begin{equation}
 \hat \phi(x) = \hat \phi_E(x) \star \hat \phi_t(x) \, ,
 \end{equation}
 where we write the field operator $\hat \phi(x)$ as the convolution of two field operators, one defining the quantum portrait of the energy, and the other one, the quantum portrait of the topology. In particular, note that the energy of the soliton only depends on $\hat \phi_E(x)$ as $E= \int dx \langle \hat \phi_E^2(x) \rangle$ with the expectation value  defined on the soliton quantum state. In addition, the topological field $\hat \phi_t(x)$ satisfies the equation
 \begin{equation}
d \hat \phi_t(x) =\pm |Q| \delta(x) \, 
\end{equation}
where $|Q|$ is given in equation (\ref{winding}) and $\pm$ correspond to soliton and anti-soliton, respectively. Thus the derivative distinguishes between solitons of opposite topological charges at the corpuscular level. Note that this equation is telling us how the topological charge is sourcing the topological component in the convolution. It is this equation that fixes the relation between the topological charge and the residue of the pole in the Fourier-transform, namely $F(\phi_t(x))(k) \sim Q/k$, i.e., the divergence in the number operator.

\section{Quantum meaning of soliton-anti-soliton interaction}
Let us now understand the corpuscular meaning of soliton-anti-soliton interaction using the language of coherent states.
The soliton-anti-soliton configuration solving the classical equations of motion is not known exactly. In general, such a configuration is not static. 
 However, we can ignore this complication and estimate the interaction between the solitons using the classical Lagrangian. The natural question raised by the previous quantum representation of the soliton lies in understanding the quantum corpuscular meaning of interaction energy {\it without }  directly using the Lagrangian information.

 Let us start with the soliton-anti-soliton classical configuration which we shall approximate by two 
static solutions separated by a finite distance, $a \, \gg \, m^{-1}$,  
 \begin{equation}
 \phi_{s,\bar s} \, =  \, \frac{m}{g}\Big( {\rm{tanh}}(m(x+a/2)) - {\rm{tanh}}(m(x-a/2) - 1\Big) \, .
 \label{solantisol}
 \end{equation}
 This can be viewed as a ``bubble" of the $\phi = + m/g$ vacuum of size $\sim a$  surrounded by  $\phi=-m/g$ asymptotic vacuum.  
  The coherent state data corresponding to this configuration is, 
  \begin{equation} 
   \alpha_k \, =   \sqrt{\pi m}  i\frac{\sqrt{k}}{g} {\rm{csch}} \Big( \frac{\pi k}{2m} \Big) \left(1- {\rm e}^{-iak}\right) \, .
\label{alphaNT}
\end{equation}
 Of course, expanding (\ref{solantisol}) in plain waves with coefficients (\ref{alphaNT}) and evaluating it in (\ref{soltensorantisol}) gives back the correct  classical
 profile after promoting the $\alpha_k$ to operators as before. The pole at $k=0$ is removed, which immediately implies that the topological charge of this configuration vanishes.

  Since the soliton-anti-soliton system has zero topological charge, the interaction energy can be estimated from the overlap 
  of the energy parts of the coherent state.  
  We can derive the exponential suppression of the soliton-anti-soliton potential by estimating the matrix element 
  \begin{equation}
  \langle E | H | E\rangle_a \, , 
 \label{matrix}
 \end{equation} 
 where  $|E\rangle_a$ stands for the energy part of the displaced anti-soliton which is the same 
 as the one of the soliton.  This matrix element can be easily estimated by taking into account  the following facts.  First, the soliton differs from the anti-soliton by the reflection of the sign. 

 Secondly, as noticed before, the translation amounts to the phase shift change $\hat{c}_k \rightarrow e^{iak} \hat{c}_k$. Moreover, the soliton and anti-soliton 
 at large separation are approximately eigenstates of the Hamiltonian. Thus,  the interaction among the solitons amounts to the overlap 
  of the two sets of coherent states with  $|c_k\rangle$ and $|e^{iak}c_k\rangle$  which from the well-known properties 
  of coherent states gives
  \begin{equation}
|\prod_k \langle c_k |e^{iak}c_k\rangle |^2\, = \,  e^{- 2 \int dk \, |c_k|^2(1-{\rm cos} (ak))}  \, . 
   \label{integral}
   \end{equation}
  Taking into account the expression for $c_k$,  we get 
  \begin{equation}
  |\prod_k \langle c_k |e^{iak}c_k\rangle |^2={\rm exp}\big(-2E_s + 32 \frac{m^3}{g^2} (am) e^{-2am}\big)\, ,
  \label{matrix1}
  \end{equation}
  where $E_s$ is given in equation (\ref{energy3}).  This expression is in agreement with the standard Lagrangian computation. 
  
  The alternative way of estimating the interaction energy is the following. 
 From the energetic point of view we can associate with the soliton and the anti-soliton two corpuscular algebras $\hat{c}_k, \hat{c}^{\dagger}_k$ and $\hat{\tilde{c}}_k,\hat{\tilde{c}}^{\dagger}_k$. In the limit of large $a$ the soliton-anti-soliton quantum state can be defined as
 \begin{equation}
 |s,\bar s\rangle = |sol\rangle \otimes |\bar{sol}\rangle
 \label{soltensorantisol}
 \end{equation}
 with $\hat{c}_k|s,\bar s\rangle = (|t_{s}\rangle \otimes c_k|E_{s}\rangle)\otimes (|t_{\bar s}\rangle \otimes |E_{\bar s}\rangle)$ and
 $\hat{\tilde c}_k|s,\bar s\rangle = (|t_{s}\rangle \otimes |E_{s}\rangle)\otimes (|t_{\bar s}\rangle \otimes \tilde c_k |E_{\bar s}\rangle)$ leading to
 \begin{equation}
 \hat{c}_k|s,\bar s\rangle = e^{iak/2} \sqrt{\frac{m^3}{\pi}}\frac{\pi k }{gm} {\rm{csch}} \Big(\frac{\pi k}{2m}\Big) |s, \bar s\rangle
 \end{equation}
 and
 \begin{equation}
 \hat{\tilde c}_k|s,\bar s\rangle = - e^{- iak/2} \sqrt{\frac{m^3}{\pi}}\frac{\pi k }{gm} {\rm{csch}} \Big(\frac{\pi k}{2m}\Big) |s, \bar s\rangle \, .
 \end{equation}
 The corpuscular Hamiltonian in this approximation becomes simply
 \begin{equation}
 H= \int d(k/m) \Big( \hat{c}^{\dagger}_k \hat{c}_k + \hat{\tilde c}^{\dagger}_k \hat{\tilde c}_k +\hat{c}^{\dagger}_k \hat{\tilde c}_k + \hat{\tilde c}^{\dagger}_k \hat{c}_k \Big) \, ,
 \label{hamiltonian2}
 \end{equation}
 where the interaction piece is manifest in the crossed terms between the two types of quanta $\hat{c}$ and $\hat{\tilde c}$. Evaluated on the state
 $|s,\bar s\rangle$ we get
 \begin{equation}
 H= 2E_s - \frac{8m^3}{g^2} (-1 + (am) {\rm{coth}} (am)) {\rm{csch}}^2(am) \, . 
 \end{equation}
 For large $a$ we obtain 
 \begin{equation}
 2E_s - 32 \frac{m^3}{g^2} (am) e^{-2am} \,, 
  \label{energy2}
 \end{equation}
 which fully agrees with (\ref{matrix1}). 
 Replacing $E_s$ by $N$ we can estimate when the interaction among the solitons becomes order one, i.e., when the effect of the interaction is equivalent to replacing $2N$ by $2N-1$. This happens for $am \sim \ln (N)$. As it is well known this fact underlies the logarithmic corrections in the soliton anti-soliton sector \cite{ZJ}. 
 
 At this point we shall briefly make few remarks on the $a$-dependence of the interaction. In contrast to an isolated soliton, for which the energy is translationally invariant, for the soliton-anti-soliton the interaction term $ -32\frac{m^3}{g^2} (am) e^{-2am}$ occurs. This term simply reflects the fact that the effective potential between the soliton and anti-soliton is attractive. The distance $a$ between them determines the interaction strength.  Due to this  
 $a$-dependent interaction the energy is no longer invariant under the change of the relative position. 
 Thus, there exists a quasi zero mode (a pseudo-Goldstone boson), which only becomes a true Goldstone mode in the limit $a\rightarrow\infty$. At the level of the corpuscular Hamiltonian (\ref{hamiltonian2}) it is easy to see that the breaking of translational invariance is due to  the interference terms $\hat{c}^{\dagger}_k \hat{\tilde c}_k$ that are mixing corpuscles of both solitons. As explained before,  shifting of the soliton profile corresponds to a phase shift of the corpuscular operators. Thus, shifting both operators in the opposite direction by an amount $b$ changes the energy by an amount $\Delta H=\int dk (e^{ik 2b} \hat{c}^{\dagger}_k \hat{\tilde c}_k+{\rm h.c.})$. In contrast shifting both profiles in the same direction will not change the energy of the system. Therefore, the interference terms explicitly break symmetry $U(1)_S\times U(1)_{\bar{S}}\rightarrow U(1)_{diag}$, resulting  in the existence of one zero mode and one quasi zero mode. 
 
 \section{Outlook} 
 
 In this paper we have displayed an attempt of uncovering the quantum origin of topology at the microscopic level. 
 For this purpose we have developed a coherent state representation of both topological and non-topological 
 solitons and confronted them with each other. 
 This construction allowed us to clearly identify the quantum origin of topological charge in terms of the singularity in the 
 occupation number of infinite wavelength corpuscles with net momentum flow in one direction. 
 After this identification, many properties of solitons, such as conservation of topological charge or false vacuum
 instability via nucleation of non-topological solitons (bubbles), nicely follow from the basic properties of the coherent states. 
 
  Our results can be straightforwardly generalized to higher co-dimension cases. In each case, the topological charge 
  is related to the singularity in occupation number of infinite wavelength quanta with the net non-zero Noether quantum number, such as, e.g., momentum or angular momentum.

  \section*{Acknowledgements}
 We thank Oscar Cata, Daniel Flassig, Slava Mukhanov and Ivo Sachs for useful discussions.
The work of G.D. was supported in part by Humboldt Foundation under Humboldt Professorship, 
ERC Advanced Grant 339169 "Selfcompletion'', by TRR 33 "The Dark
Universe" and by by the DFG cluster of excellence "Origin and Structure of the Universe". 
The work of C.G. was supported in part by Humboldt Foundation and by Grants: FPA 2009-07908, CPAN (CSD2007-00042) and by the ERC Advanced Grant 339169 "Selfcompletion'' .
The work of L.G. and T.R. was supported by the International Max Planck Research School on Elementary
Particle Physics.

\end{document}